\documentclass[twocolumn,superscriptaddress,pra]{revtex4-2}
\usepackage[utf8]{inputenc}
\pdfoutput=1
\usepackage{hyperref}
\usepackage{orcidlink}

\pdfoutput=1
\usepackage[english]{babel}
\usepackage[T1]{fontenc}

\usepackage{amsmath}
\usepackage{bbold}

 \def\ket#1{|#1\rangle}

\begin{document}

\title{Towards minimal self-testing of qubit states and measurements in prepare-and-measure scenarios}%
\thanks{New J. Phys. 26, 063012 (2024), \url{https://doi.org/10.1088/1367-2630/ad4e5c}. Original Content from this work may be used under the terms of the Creative Commons Attribution 4.0 licence. Any further distribution of this work must maintain attribution to the authors and the title of the work, journal citation and DOI.}%

\author{G\'abor Dr\'otos\,\orcidlink{0000-0002-0900-5188}}
\affiliation{MTA Atomki Lend\"ulet Quantum Correlations Research Group, HUN-REN Institute for Nuclear Research, P.O. Box 51, H-4001 Debrecen, Hungary}

\author{K\'aroly F. P\'al}
\affiliation{HUN-REN Institute for Nuclear Research, P.O. Box 51, H-4001 Debrecen, Hungary}

\author{Abdelmalek Taoutioui\,\orcidlink{0000-0002-4943-5529}}
\affiliation{MTA Atomki Lend\"ulet Quantum Correlations Research Group, HUN-REN Institute for Nuclear Research, P.O. Box 51, H-4001 Debrecen, Hungary}

\author{Tam\'as V\'ertesi\,\orcidlink{0000-0003-4437-9414}}
\email{tvertesi@atomki.hu}
\affiliation{MTA Atomki Lend\"ulet Quantum Correlations Research Group, HUN-REN Institute for Nuclear Research, P.O. Box 51, H-4001 Debrecen, Hungary}

\begin{abstract}
Self-testing is a promising approach to certifying quantum states or measurements. Originally, it relied solely on the outcome statistics of the measurements involved in a device-independent (DI) setup. Extra physical assumptions about the system make the setup semi-DI. In the latter approach, we consider a prepare-and-measure scenario in which the dimension of the mediating particle is assumed to be two. In a setup involving four (three) preparations and three (two) projective measurements in addition to the target, we exemplify how to self-test any four- (three-) outcome extremal positive operator-valued measure using a linear witness. One of our constructions also achieves self-testing of any number of states with the help of as many projective measurements as the dimensionality of the space spanned by the corresponding Bloch vectors. These constructions are conjectured to be minimal in terms of the number of preparations and measurements required. In addition, we implement one of our prepare-and-measure constructions on IBM and IonQ quantum processors and certify the existence of a complex qubit Hilbert space based on the data obtained from these experiments. 
\end{abstract}

\maketitle

\section{Introduction}

Quantum technological applications need certified devices. Ideally, certification does not rely on any assumption about the internal functioning of the devices involved in the protocol, in which case it is called device-independent (DI)~\cite{acin2007device,scarani2012device}. In particular, if the characterization of the preparation and measurement apparatuses is solely based on observed measurement statistics, we speak of self-testing in the original sense, as proposed by Mayers and Yao~\cite{mayers2004}. Such a self-testing is made possible by exploiting the Bell nonlocal property of multipartite quantum correlations~\cite{bell1964,brunner2014bell}.

However, fully DI certification is difficult to implement. This has led to the appearance of the so-called semi-device-independent (semi-DI) approach, where certain extra physical assumptions are made about the devices~\cite{pawlowski2011semi}. Particular assumptions refer to bounds on overlap~\cite{brask2017megahertz,tebyanian2021}, mean energy~\cite{van2017semi}, entropy of prepared states~\cite{chaves2015device}, or may be based on the response of physical systems to spatial rotations~\cite{jones2022}. In addition, still in the semi-DI framework, all measurement devices can be considered as fully trusted~\cite{vallone2014, marangon2017} or only some of them can be considered as trusted~\cite{sarkar2022,sarkar2022}. In the standard prepare-and-measure (PM) scenario, which will form the framework for self-testing in our paper, the most widely accepted assumption is an upper bound on the dimension of the communicated quantum system (see, e.g.,~\cite{ambainis2008quantum,woodhead2015,lunghi2015,tavakoli2018self,farkas2019self,mironowicz2019,tavakoli2020self,deGois2021,divianszky2022,alves2023PRA,farkas2023}).

In the PM scenario, there are two parties, say Alice and Bob. Upon receiving an input $x$, Alice prepares a state $\rho_x$ and sends it to Bob, who performs a measurement on this state. The choice of his measurement setting is determined by his input $y$, and the outcome is labeled $b$. A self-test is performed if using the statistics of the outcomes of the measurements, $P(b|xy)$, one can prove that the states and the measurement settings are the ones they are claimed to be (up to some symmetry transformation). Within the PM scenario, only semi-DI tests can be performed~\cite{gallego2010device}. In fact, suppose that the dimension $d$ of the mediating particle can be arbitrarily large to perfectly encode the input $x$. Then Alice simply sends $x$ encoded into a $d$ dimensional classical message to Bob, who has all the information about both Alice's and his own input and can therefore simulate any possible conditional probability $P(b|xy)$.

Our aim in this paper is to provide possibly minimal PM setups, in terms of the number of involved states and measurement settings, for self-testing any member of a rather general class of measurements, that of extremal four- or three-outcome positive operator-valued measures (POVMs) defined over a two-dimensional Hilbert space; in the most general form, this will be achieved by constructing possibly minimal PM setups for self-testing an arbitrary number of pure states of the same Hilbert space.

An $O$-outcome POVM is defined by its $O$ elements, which are positive semidefinite operators whose sum is the identity operator~\cite{peres1997book,nielsen2002}. A POVM is said to be extremal if it cannot be written as a convex combination of other POVMs. If a POVM has more outcomes than the dimensionality of the corresponding Hilbert space, it is necessarily non-projective, that is, its elements are not orthogonal projectors, which is a constraint defining a traditional von Neumann measurement~\cite{holevo2003}. Non-projective POVMs provide benefits through enhanced entanglement detection~\cite{shang2018,bae2019} and state discrimination~\cite{dieks1988,peres1988}. Among their applications, we can list quantum protocols such as quantum coin flipping, quantum money, and quantum cryptography~\cite{bennett1992}.

Existing literature on self-testing non-projective POVMs mostly concerns a paradigmatic but specific measurement, the so-called symmetric informationally complete (SIC) POVM~\cite{renes2004,fuchs2017} defined over a Hilbert space of dimension two~\cite{mironowicz2019,tavakoli2020self} or arbitrary dimension~\cite{tavakoli2019PRL}. However, compounds of SIC POVMs~\cite{Tavakoli2020b} and the so-called semi-SIC POVMs~\cite{Geng2021,drotos} have also been self-tested.

Most recently, the task of self-testing arbitrary ensembles of pure states and arbitrary sets of extremal POVMs in PM scenarios of arbitrary Hilbert space dimension has already been accomplished in Ref.~\cite{navascues2023}. However, this work does not, in general, rely on a minimal number of preparations and measurement settings. We also have to mention Ref.~\cite{sarkar2023}, where arbitrary rank-one extremal measurements are self-tested within the so-called one-sided DI framework~\cite{supic2016}. In addition to the assumption that the measurement device is trusted, another important difference from our work is the use of EPR steering~\cite{Schrodinger1935,Wiseman2007} as a resource.

For our construction, we will adopt the scheme introduced in Ref.~\cite{tavakoli2020self} for self-testing non-projective POVMs in a PM scenario where Alice prepares qubits, that is, states in a two-dimensional Hilbert space. This scheme is based on choosing the preparations of Alice and a set of ``auxiliary'' measurement settings on Bob's side, all of them projective, such that the maximum value of a carefully constructed linear witness (that is, a function of the probabilities $P(b|xy)$ associated with each measurement outcome $b$ conditioned on the inputs $x$ and $y$) self-tests the target POVM which enters as the ``main'' measurement of Bob. Moreover, we will closely follow the particular path developed in our preceding study~\cite{drotos} in which the members of a specific class of non-projective measurements, the semi-SIC POVMs~\cite{Geng2021,zhao2024PLA}, are self-tested, and heavily build on certain results obtained in that context. We emphasize that our self-testing methodology is fully analytic and does not address the effect of noise. However, in addition to self-testing, we present a method to certify a complex qubit Hilbert space in the PM scenario in the presence of experimental noise.

The paper is organized as follows. We start by defining the setup and laying down the basic ideas in Section~\ref{sec:preliminaries}. We continue in Section~\ref{sec:general} by providing universal considerations with regard to the self-testing of the prepared states for later use. In Section~\ref{sec:four-outcome}, we construct a potentially minimal setup for self-testing four-outcome POVMs. A more general construction is presented in Section~\ref{sec:anynumber}, which covers the self-testing of POVMs with four, but also three outcomes and any number of prepared states. We conclude in Section~\ref{sec:conclusion}. Appendix~\ref{sec:example} is devoted to an illustrative example, while Appendix~\ref{sec:4x6} presents an alternative method for self-testing four-outcome POVMs, although not relying on a minimal setup, as a special case of the results obtained in Ref.~\cite{navascues2023}. In Appendix~\ref{sec:implementing}, we construct a particular complex qubit witness in the prepare-and-measure setup, which we implement on IBM and IonQ quantum processors thereby certifying a complex qubit Hilbert space from observed data. The details of the implementation of the setup on quantum processors can be found in Appendix~\ref{sec:qcircuit}.

\section{Preliminaries}\label{sec:preliminaries}

In the qubit space, the element corresponding to outcome $b$, $b\in\{1,\ldots,O\}$, of an $O$-outcome POVM can be written in terms of its Bloch vector $\vec n_b$ as:
\begin{equation}
M_b=\lambda_b(\mathbb{1}+\vec n_b\cdot\vec\sigma),
\label{eq:povmgen}
\end{equation}
where $|\vec n_b|\leq 1$, $\lambda_b>0$ (these conditions ensure positivity), $\sum_b\lambda_b=1$ and $\sum_b\lambda_b\vec n_b=0$ (hence $\sum_b M_b=\mathbb{1}$), and the components of $\vec\sigma$ are the Pauli matrices. Note that only extremal POVMs can be self-tested with a linear witness (see, e.g., Ref.~\cite{navascues2023}). The maximum number of outcomes for an extremal POVM in the qubit space is four. Such a POVM is extremal if and only if its four Bloch vectors are unit vectors that span the three-dimensional space (actually, due to the condition $\sum_b\lambda_b\vec n_b=0$, any three of them do). In the case of three outcomes, the Bloch vectors must span a two-dimensional subspace for extremality. In the case of two outcomes, the extremal POVMs are projective measurements. We will deal with self-tests of four- and three-outcome cases.

To perform the self-test, we consider the PM setup proposed in Ref.~\cite{tavakoli2020self} and described as follows. Let the number of input values for Alice be denoted by $M_m$ and be equal to the number of outcomes of the POVM, that is, $M_m=O$. (Under certain circumstances, we will deviate from this equality, but it does not affect the relevant considerations.)
Let the number of input values for Bob, that is, the number of measurement settings, be $M_v+1$. Let the first $M_v$ settings be two-outcome projective measurements with outcomes labeled by $b$, $b\in\{0,1\}$, while the $(M_v+1)$th one the POVM to be self-tested.

Let us also adopt the assumption that Alice prepares qubits, that is, states of the form
\begin{equation}
\rho_x=\frac{\mathbb{1}+\vec m_x\cdot\vec\sigma}{2},
\label{eq:rho}
\end{equation}
where the Bloch vectors $\vec m_x$, $|\vec m_x| \leq 1$, provide a unique characterization.

The linear witness introduced in Ref.~\cite{tavakoli2020self} reads as:
\begin{equation}
W'=W-k\sum_{b=1}^{O}P(b=x|x(M_v+1)),
\label{eq:witnesswithprobfull}
\end{equation}
with
\begin{equation}
W=\sum_{x=1}^{M_m}\sum_{y=1}^{M_v}\sum_{b=0}^{1} w_{xy}^{(b)}P(b|xy),
\label{eq:witnesswithprob}
\end{equation}
where $P(b|xy)$ is the conditional probability of Bob getting outcome $b$ given $x$ and $y$, while $k$ is a positive number.

For the self-test, the prepared states must be pure, $|\vec m_x| = 1$, and their Bloch vectors must be anti-aligned with the ones of the POVM to be self-tested such that $\vec m_x=-\vec n_b$ for $b=x$. Furthermore, the matrices of elements $w_{xy}^{(b)}$ and the $M_v$ projective binary measurements of Bob must be chosen such that the formula given in Eq.~(\ref{eq:witnesswithprob}) should be appropriate to self-test these binary measurements and the states with the previously mentioned Bloch vectors; that is, $W$ should be maximal for these states and measurements, and only for them. With these choices, the expression in Eq.~(\ref{eq:witnesswithprobfull}) is appropriate to self-test the non-projective POVM [the ($M_v+1$)th measurement], too; the optimum value of the second term is zero, and it takes this value only if $\vec n_x=-\vec m_b$ for $x=b$. The Bloch vectors $\vec n_b$ uniquely define the POVM if it is extremal, because then there is only one way to combine the zero vector from them. It is not so when, for example, there are four vectors in the same plane, which corresponds to a non-extremal POVM.

\section{General considerations on self-testing the prepared states}\label{sec:general}

We address here the self-testing of the prepared states without specifying the form of the matrices of elements $w_{xy}^{(b)}$.

The witness for self-testing the prepared states is given by Eq.~(\ref{eq:witnesswithprob}). Using $P(0|xy)+P(1|xy)=1$, this equation can be rewritten as:
\begin{multline}
W=\frac{1}{2}\sum_{x=1}^{M_m}\sum_{y=1}^{M_v}\left(w_{xy}^{(0)}+w_{xy}^{(1)}\right)\\+\frac{1}{2}\sum_{x=1}^{M_m}\sum_{y=1}^{M_v}\left(w_{xy}^{(0)}-w_{xy}^{(1)}\right)\left[P(0|xy)-P(1|xy)\right].
\label{eq:witnesswithprob2}    
\end{multline}
The first term is an irrelevant constant, we can get rid of it without loss of generality by choosing $w_{xy}\equiv w_{xy}^{(0)}=-w_{xy}^{(1)}$ (or, alternatively, by dropping the constant and introducing the notation $w_{xy}\equiv\left(w_{xy}^{(0)}-w_{xy}^{(1)}\right)/2$). Doing that, one gets
\begin{equation}
W=\sum_{x=1}^{M_m}\sum_{y=1}^{M_v}w_{xy}\left[P(0|xy)-P(1|xy)\right].
\label{eq:witnesswithprob3}
\end{equation}
We note that $P(0|xy)-P(1|xy)$ is the expectation value of an observable that takes on the values $+1$ and $-1$ upon the measurement outcomes $b=0$ and $b=1$, respectively.

The conditional probability of getting outcome $b$ when a POVM measurement is performed on a qubit state $\rho_x$ is the following:
\begin{equation}
P(b|xy)=\mathrm{Tr}(\rho_xM_{b|y}),
\label{eq:condprob}
\end{equation}
where $M_{b|y}$ is the POVM element of measurement $y$ corresponding to outcome $b$.
A qubit state $\rho_x$ is characterized by its Bloch vector $\vec m_x$ according to Eq.~(\ref{eq:rho}). The most general form of POVM elements $M_{b|y}$, $b\in\{0,1\}$, of a two-outcome measurement setting $y$ can be written as:
\begin{equation}
M_{b|y}=\left[1+(-1)^b\mu_y\right]\frac{\mathbb{1}+(-1)^b\vec v_y^{(b)}\cdot\vec\sigma}{2},
\label{eq:povmelement}
\end{equation}
with the Bloch vectors satisfying
\begin{equation}
(1+\mu_y)\vec v_y^{(0)}=(1-\mu_y)\vec v_y^{(1)}\equiv(1-|\mu_y|)\vec v_y,
\label{eq:povmconstr}
\end{equation}
where $|\vec v_y^{(b)}|\leq 1$, and $\vec v_y$ is defined as the longer one of $\vec v_y^{(0)}$ and $\vec v_y^{(1)}$. The values $\mu_y=0$ and $\mu_y=\pm 1$ correspond to a genuine von Neumann measurement and the degenerate measurements (fixed outcome), respectively. From the above equations, the difference of the two POVM elements is:
\begin{equation}
M_{0|y}-M_{1|y}=\mu_y\mathbb{1}+(1-|\mu_y|)\vec v_y\cdot\vec\sigma.
\label{eq:diffpovmelem}
\end{equation}
Using Eqs.~(\ref{eq:rho},\ref{eq:witnesswithprob3},\ref{eq:condprob}), and Eq.~(\ref{eq:diffpovmelem}) one can show that:
\begin{equation}
W=\sum_{y=1}^{M_v}\sum_{x=1}^{M_m}w_{xy}[\mu_y+(1-|\mu_y|)\vec m_x\cdot\vec v_y].
\label{eq:Wwithdeg}
\end{equation}

To obtain the maximum value of $W$ (which we denote as $Q\equiv W_{\mathbb{C}^2}$ with reference to being a maximum attainable quantum mechanically or, more specifically, by means of complex qubits), it is clear that if $\sum_xw_{xy}\vec m_x\cdot\vec v_y>|\sum_x w_{xy}|$, then $\mu_y=0$, therefore, a genuine von Neumann measurement should be performed, while if $\sum_xw_{xy}\vec m_x\cdot\vec v_y<|\sum_x w_{xy}|$, the optimal measurement is degenerate.
In the latter case, the measurement is useless and $w_{xy}$ is inappropriate for the self-test. If $\sum_xw_{xy}=0$, that is, the sum of the elements of the $y$th column is zero, then the contribution of the degenerate measurement is zero. However, it follows from Appendix~B of Ref.~\cite{drotos} that if all $\vec m_x$ are optimal and not all elements of the $y$th column of $w_{xy}$ are zero, then the contribution of the genuine two-outcome von Neumann measurement corresponding to the optimal $\vec v_y$ is positive. Therefore, if $w_{xy}$ is such that the sum of its elements in all columns is zero, the optimum does not contain any degenerate measurement. 

Restricting ourselves only to genuine two-outcome measurements, that is, taking $\mu_y=0$ in Eq.~(\ref{eq:Wwithdeg}), the expression to maximize in terms of unit vectors $\vec m_x$ and $\vec v_y$ is:
\begin{equation}
W=\sum_{x=1}^{M_m}\sum_{y=1}^{M_v} w_{xy}\vec m_x\cdot\vec v_y=\sum_{x=1}^{M_m}\vec m_x\cdot\sum_{y=1}^{M_v}w_{xy}\vec v_y.
\label{eq:Wfor3X4}
\end{equation}

We will follow the procedure presented in \cite{drotos}. It is easy to see that the optimum choice for $\vec m_x$ if vectors $\vec v_y$ are given is:
\begin{equation}
\vec m_x=\frac{\vec u_x}{|\vec u_x|},
\label{eq:optimx}
\end{equation}
where
\begin{equation}
\vec u_x\equiv\sum_{y=1}^{M_v}w_{xy}\vec v_y,
\label{eq:vecux}
\end{equation}
if $|\vec u_x|\neq 0$, which condition is always true for all $x$ if $\vec v_y$ is optimal and each row of $w_{xy}$ contains at least one non-zero element, as it is proven in Appendix~B of Ref.~\cite{drotos}. Using Eqs.~(\ref{eq:Wfor3X4}-\ref{eq:vecux}) it is easy to see that the value of $W$ with the optimal choice of $\vec m_x$ for given $\vec v_y$ can be written as:
\begin{equation}
Q_v=\sum_{x=1}^{M_m}|\vec u_x|.
\label{eq:Qv}
\end{equation}
This expression is valid even if $\vec v_y$ is such that $|\vec u_x|=0$ for some $x$ (in this case, any $\vec m_x$ is optimal for the given $x$). Using Eq.~(\ref{eq:vecux}), $|\vec u_x|$ can be written as:
\begin{equation}
|\vec u_x|=\sqrt{\sum_{y=1}^{M_v}w_{xy}^2+\sum_{y=1}^{M_v-1}\sum_{y'=y+1}^{M_v}w_{xy}w_{xy'}\gamma_{yy'}},
\label{eq:uxwithgamma}
\end{equation}
where $\gamma_{yy'}\equiv\vec v_y\cdot\vec v_{y'}$ is an element of the Gram matrix of the set $\vec v_y$ (as $\vec v_y$ are normalized, the diagonal elements of the Gram matrix are $1$). We need to find the maximum of $Q_v$ in terms of $\vec v_y$. The set of $\vec v_y$ is never unique, it may only be determined up to a global orthogonal transformation. It is the Gram matrix of this set of vectors that can be uniquely defined. It has been proven in Appendix~C of Ref.~\cite{drotos} that if the number $M_v$ of the binary measurement settings is equal to the dimensionality of the space, and the derivatives of $Q_v=\sum|u_x|$ with respect to $\gamma_{yy'}$, $y<y'$, vanish somewhere, then the value of $Q_v$ takes its global maximum there, provided that these $\gamma_{yy'}$ values correspond to a legitimate Gram matrix (that is, the symmetric matrix whose diagonal elements are $1$ and the $yy'$th element is $\gamma_{yy'}$ is positive semidefinite). The Gram matrix is uniquely defined if the condition is valid at a single point. 

Recall that we are, in fact, interested in constructing a witness matrix $w$ and choosing corresponding vectors $\vec v_y$ that make self-testing possible for given $\vec m_x$. Therefore, even if the optimal $\vec v_y$ and $\vec m_x$ are determined for a given $w_{xy}$ only up to a global orthogonal transformation, we will need to specify $\vec v_y$ in terms of a particularly oriented set of $\vec m_x$.

\section{Self-testing four-outcome POVMs}\label{sec:four-outcome}

For the specific task of self-testing four-outcome POVMs, $O = 4$ (which implies $M_m = 4$ unless otherwise stated), we fix the number of binary measurement settings to $M_v = 3$, and construct an appropriate witness matrix, in terms of which the appropriate Bloch vectors can also be determined.

Let us consider a witness matrix of the following form:
\begin{equation}
w=\left(\begin{array}{ccc}
  \hphantom{-}p_1q_1&\hphantom{-}p_1q_2&\hphantom{-}p_1q_3\\
  \hphantom{-}p_2q_1&-p_2q_2&-p_2q_3\\
  -p_3q_1&\hphantom{-}p_3q_2&-p_3q_3\\  	
  -p_4q_1&-p_4q_2&\hphantom{-}p_4q_3\\
  \end{array}\right),
\label{eq:w4X3}
\end{equation}
with
\begin{equation}
\sum_{x=1}^4p_x^2=\sum_{y=1}^3q_y^2=1.
\label{eq:pqnorm}
\end{equation}

From Eqs.~\ref{eq:uxwithgamma} and \ref{eq:w4X3} it follows that
\begin{align}
|\vec u_1|=&|p_1|\sqrt{1+2(q_1q_2\gamma_{12}+q_1q_3\gamma_{13}+q_2q_3\gamma_{23})},\nonumber\\
|\vec u_2|=&|p_2|\sqrt{1+2(-q_1q_2\gamma_{12}-q_1q_3\gamma_{13}+q_2q_3\gamma_{23})},\nonumber\\
|\vec u_3|=&|p_3|\sqrt{1+2(-q_1q_2\gamma_{12}+q_1q_3\gamma_{13}-q_2q_3\gamma_{23})},\nonumber\\
|\vec u_4|=&|p_4|\sqrt{1+2(q_1q_2\gamma_{12}-q_1q_3\gamma_{13}-q_2q_3\gamma_{23})}.
\label{eq:vecuy}
\end{align}
From Eqs.~\ref{eq:Qv} and \ref{eq:vecuy} it is easy to see that if $q_y\neq 0$ for all $y$, the condition for the partial derivatives can be written as:
\begin{align}
\frac{p_1^2}{|\vec u_1|}-\frac{p_2^2}{|\vec u_2|}-\frac{p_3^2}{|\vec u_3|}+\frac{p_4^2}{|\vec u_4|}&=0,\nonumber\\
\frac{p_1^2}{|\vec u_1|}-\frac{p_2^2}{|\vec u_2|}+\frac{p_3^2}{|\vec u_3|}-\frac{p_4^2}{|\vec u_4|}&=0,\nonumber\\
\frac{p_1^2}{|\vec u_1|}+\frac{p_2^2}{|\vec u_2|}-\frac{p_3^2}{|\vec u_3|}-\frac{p_4^2}{|\vec u_4|}&=0.
\label{eq:deriv}
\end{align}
This means that
\begin{equation}
\frac{|\vec u_1|}{p_1^2}=\frac{|\vec u_2|}{p_2^2}=\frac{|\vec u_3|}{p_3^2}=\frac{|\vec u_4|}{p_4^2}\equiv u.
\label{eq:eqcond}
\end{equation}
Substituting Eq.~(\ref{eq:vecuy}) into the square of the equations above we get:
\begin{align}
p_1^2u^2&=1+2(q_1q_2\gamma_{12}+q_1q_3\gamma_{13}+q_2q_3\gamma_{23}),\nonumber\\
p_2^2u^2&=1+2(-q_1q_2\gamma_{12}-q_1q_3\gamma_{13}+q_2q_3\gamma_{23}),\nonumber\\
p_3^2u^2&=1+2(-q_1q_2\gamma_{12}+q_1q_3\gamma_{13}-q_2q_3\gamma_{23}),\nonumber\\
p_4^2u^2&=1+2(q_1q_2\gamma_{12}-q_1q_3\gamma_{13}-q_2q_3\gamma_{23}).
\label{eq:p2u2}
\end{align}
From the sum of the equations above it follows that $u=2$. Then from Eq.~(\ref{eq:eqcond}) we get that $|\vec u_x|=2p_x^2$, and the optimum value of $Q_v$, and consequently $W$ is $Q=2$. Here we also used Eq.~(\ref{eq:pqnorm}). We get the optimum values of $\gamma_{12}$, $\gamma_{13}$ and $\gamma_{23}$ if we add the fourth, the third and the second line of Eq.~(\ref{eq:p2u2}) to the first one, respectively:
\begin{align}
\gamma_{12}&=\frac{p_1^2+p_4^2-1/2}{q_1q_2},\nonumber\\
\gamma_{13}&=\frac{p_1^2+p_3^2-1/2}{q_1q_3},\nonumber\\
\gamma_{23}&=\frac{p_1^2+p_2^2-1/2}{q_2q_3}.
\label{eq:optimgamma}
\end{align}
If the equations above together with $\gamma_{yy'}=\gamma_{y'y}$ and $\gamma_{yy}=1$ result in a legitimate Gram matrix, it corresponds to the unique solution of the problem, and it determines the optimum $\vec v_y$ up to an orthogonal transformation. For an optimum set $\vec v_y$ the optimum $\vec m_x$ can be written as:
\begin{align}
\vec m_1&=\frac{q_1\vec v_1+q_2\vec v_2+q_3\vec v_3}{2p_1},\nonumber\\
\vec m_2&=\frac{q_1\vec v_1-q_2\vec v_2-q_3\vec v_3}{2p_2},\nonumber\\
\vec m_3&=\frac{-q_1\vec v_1+q_2\vec v_2-q_3\vec v_3}{2p_3},\nonumber\\
\vec m_4&=\frac{-q_1\vec v_1-q_2\vec v_2+q_3\vec v_3}{2p_4},
\label{eq:mvec}
\end{align}
which follows from Eqs.~(\ref{eq:optimx}, \ref{eq:vecux}) and Eq.~(\ref{eq:w4X3}), and from $|\vec u_x|=2p_x^2$.

However, our objective is not to find the vectors that maximize $W$ for the given matrix $w$, but to find the appropriate matrix $w$ if the $\vec m_x$ vectors are given, which, multiplied by minus one, define the four-outcome POVM to be self-tested. From Eq.~(\ref{eq:mvec}) it follows that
\begin{equation}
\sum_x p_x\vec m_x=0.
\label{eq:pdef}
\end{equation}
If $\vec m_x$ span the three-dimensional space, this, together with the normalization condition, uniquely defines the parameters $p_x$ in Eq.~(\ref{eq:w4X3}) up to an overall sign. In terms of the parameters $q_y$ concerned, from the first two lines of Eq.~(\ref{eq:mvec}) we get $p_1\vec m_1+p_2\vec m_2=q_1\vec v_1$. Let us choose the sign of $q_1$ positive (choosing it negative would simply reverse the direction of the optimal $\vec v_1$). We can get the values $q_2$ and $q_3$ analogously, the result is:
\begin{align}
q_1&=|p_1\vec m_1+p_2\vec m_2|,\nonumber\\
q_2&=|p_1\vec m_1+p_3\vec m_3|,\nonumber\\
q_3&=|p_1\vec m_1+p_4\vec m_4|.
\label{eq:qvalues}
\end{align}
It can be proven that the $q_y$ values calculated this way satisfy the normalization condition, Eq.~(\ref{eq:pqnorm}).

We follow with a remark on $p_x$. The linear combination of the Bloch vectors $\vec n_b$ with the coefficients appearing in the expression for the POVM elements, $\lambda_b$, gives the null vector, all those coefficients are positive, and their sum is one. As $\vec m_x=-\vec n_b$ for $x=b$, it then follows from Eq.~(\ref{eq:pdef}) that those $p_x$ are just proportional to $\lambda_b$ for $x=b$, and they are all positive. They should just be renormalized. The vectors $\vec n_b$ define the coefficients $\lambda_b$ uniquely if they span a three-dimensional space. If not, an infinite number of POVMs can be defined with the same Bloch vectors; therefore, such POVMs cannot be self-tested, which is what is expected, as such POVMs are not extremal.

The $w$ matrix constructed here is such that the maximum of Eq.~(\ref{eq:Wfor3X4}) is given by the Bloch vectors of the states to be self-tested. Moreover, the solution is unique up to orthogonal transformations. This is actually true for any set of four qubit states, including those that do not define a POVM (not all $p_x$ are positive). However, depending on the set of vectors, it may well happen that an even larger value can be obtained if degenerate measurements are allowed, as in Eq.~(\ref{eq:Wwithdeg}). This must always be checked, because if this is the case, the $w$ matrix is inappropriate for the self-test. A way to solve this problem is to add four additional rows to $w$ containing the same elements as the first four rows but with opposite signs. The set of states with Bloch vectors of opposite directions should also be included in the self-test, therefore, Alice has to be able to prepare twice as many states. As the sum of the elements of each column of $w$ will obviously be zero now, the degenerate measurements will always be inferior, making the self-test feasible.

Once an appropriate $w$ matrix has been fixed as above, one also has to determine, in terms of $\vec m_x$, the corresponding Bloch vectors $\vec v_y$ characterizing the projective measurements that lead to the maximum value of the witness. This could be done by inverting Eqs.~(\ref{eq:mvec}). However, in analogy with the proof presented in Appendix~B of Ref.~\cite{drotos}, it can be shown that the optimum $\vec v_y$ can be expressed in terms of $w_{xy}$ and $\vec m_x$ as:
\begin{equation}
\vec v_y=\frac{\sum_{x=1}^{4}w_{xy}\vec m_x}{|\sum_{x=1}^{4}w_{xy}\vec m_x|}.
\label{eq:optimv}
\end{equation}

We will now briefly discuss two special configurations of the vectors $\vec m_x$. First, if they are in the same plane, they do not define $p_x$ uniquely. Then the optimum of Eq.~(\ref{eq:Wfor3X4}) will be given by the same set of vectors $\vec m_x$ using any of the possible choices for $p_x$, so that the states corresponding to $\vec m_x$ can still be self-tested, even though the corresponding POVM will not be uniquely defined, as discussed earlier. Another interesting arrangement is three coplanar vectors with the fourth one orthogonal to them. In this case the fourth row of $w$ becomes zero. If we omit this row and the fourth vector, then the resulting $3\times 3$ matrix is appropriate to self-test three coplanar vectors, provided that the three genuine two-outcome measurements given by the optimal $\vec v_y$ vectors are the most advantageous. Otherwise, the number of rows and states can be doubled, as before. In this way, it is possible to self-test three-outcome extremal POVM measurements with the help of three additional measurement settings. We will see that with a different method, the same is possible with just two settings.

\section{Self-testing any number of states and POVMs of four and three outcomes}\label{sec:anynumber}

Now we will discuss how to self-test a set of qubit states of arbitrary number in a possibly minimal PM setup. The corresponding witness $W$ is defined by Eq.~(\ref{eq:witnesswithprob}), and we will rely on the corresponding considerations of Section~\ref{sec:general}. Whenever the Bloch vectors $\vec m_x$ of the states uniquely define a POVM according to Eq.~(\ref{eq:povmgen}) through $\vec n_b=-\vec m_x$ for $b=x$, that POVM can also be self-tested by choosing $\vec m_x=-\vec n_b$ for $x=b$ and extending the witness according to Eq.~(\ref{eq:witnesswithprobfull}). We will use this fact to show how to self-test three-outcome POVMs in a possibly minimal setup beyond four-outcome POVMs.

In particular, only two projective measurements are required in this general framework if the Bloch vectors are coplanar, which is the case for three-outcome POVMs. The number of projective measurements remains three for Bloch vectors spanning the three-dimensional space, such as for extremal four-outcome POVMs. In general, the number $M_v$ of projective measurements must be equal to the dimensionality of the space spanned by the Bloch vectors $\vec m_x$ of the $M_m$ states to be self-tested.

In any case, the appropriate projective measurements are orthogonal to each other. There is also considerable freedom in choosing the $w$ matrix, which may be used to make sure that all degenerate measurements are inferior to the genuine two-outcome ones so that $w$ is appropriate indeed, and possibly to optimize properties of the self-test.

Let the elements of the $w$ matrix be
\begin{equation}
w_{xy}=r_x\mu_{xy},
\label{eq:wanysize}
\end{equation}
where
\begin{equation}
\sum_{y=1}^{M_v}\mu_{xy}^2=1.
\label{eq:munorm}
\end{equation}
From Eq.~(\ref{eq:uxwithgamma}) it follows that for the present choice of the $w$ matrix
\begin{equation}
|\vec u_x|=|r_x|\sqrt{1+2\sum_{y=1}^{M_v-1}\sum_{y'=y+1}^{M_v}\mu_{xy}\mu_{xy'}\gamma_{yy'}},
\label{eq:absuanysize}
\end{equation}
where $\gamma_{yy'}\equiv\vec v_y\cdot\vec v_{y'}$. The conditions for the partial derivatives to maximize $Q_v=\sum_x|\vec u_x|$, as defined in Eq.~(\ref{eq:Qv}), can be written as
\begin{multline}
\frac{\partial Q_v}{\partial\gamma_{y_1y_2}}=\sum_{x=1}^{M_m}\frac{|r_x|\mu_{xy_1}\mu_{xy_2}}{\sqrt{1+2\sum_{y=1}^{M_v-1}\sum_{y'=y+1}^{M_v}\mu_{xy}\mu_{xy'}\gamma_{yy'}}}\\=0.
\label{eq:derconanysize}
\end{multline}

Let the parameters of the $w$ matrix satisfy
\begin{equation}
\sum_{x=1}^{M_m}|r_x|\mu_{xy_1}\mu_{xy_2}=0
\label{eq:rmuanysize}
\end{equation}
for all $y_1<y_2$. Then the conditions for the derivatives in Eq.~(\ref{eq:derconanysize}) are satisfied if all $\gamma_{yy'}=0$, that is, orthogonal Bloch vectors for the measurement settings are an optimal choice (it maximizes $Q_v$). In this case, $|\vec u_x|=|r_x|$ from Eq.~(\ref{eq:absuanysize}), and $Q_v=\sum_x|r_x|$. Let all $r_x$ be positive. Then it follows, according to Eq.~(\ref{eq:optimx}), that
\begin{equation}
\vec m_x=\frac{\vec u_x}{|\vec u_x|}=\sum_{y=1}^{M_v}\mu_{xy}\vec v_y,
\label{eq:optimxanysize}
\end{equation}
that is, the $y$th component of $\vec m_x$ in the coordinate system defined by the orthonormal set of the measurement vectors is nothing else than $\mu_{xy}$ (hence the equality between the number $M_v$ of measurement settings and the dimensionality of the space spanned by the $M_m$ vectors $\vec m_x$). A matrix $w$ defined through such parameters $\mu_{xy}$ may only be appropriate for the self-test if $\gamma_{yy'}=0$ is the only optimum, we will discuss this question later.

To self-test a given set of $M_m$ qubit states characterized by Bloch vectors $\vec m_x$, we have to choose the parameters $\mu_{xy}$ to be the components of $\vec m_x$ in a coordinate system. However, only some frames are appropriate. For example, Eqs.~(\ref{eq:rmuanysize}) cannot be satisfied if all $\mu_{xy_1}\mu_{xy_2}$ have the same sign for all $x$ for any $y_1<y_2$. If $\mu_{xy_1}$ and $\mu_{xy_2}$ are components of $\vec m_x$ in a frame, then $\mu_{xy_1}\mu_{xy_2}$ is an off-diagonal component of the outer product of $\vec m_x$ with itself in the same frame. This outer product $\vec m_x\otimes\vec m_x$ is the projection operator corresponding to $\vec m_x$. Then Eq.~(\ref{eq:rmuanysize}) means that the matrix of the operator $\sum_xr_x\vec m_x\otimes\vec m_x$ is diagonal. Therefore, we can construct a $w$ choosing $r_x$ to be some positive numbers, and $\mu_{xy}$ to be the components of the set $\vec m_x$ in the coordinate system in which $\sum_x r_x\vec m_x\otimes\vec m_x$ is diagonal. Such a frame exists as the operator is positive semidefinite.

For $w$ to be appropriate for the self-test it is necessary that all degenerate measurements give smaller $W$ values than the optimal non-degenerate ones. This is not true for all choices of $r_x$. For vectors $\vec m_x$ that define POVM measurements as discussed at the beginning of the section, one can always ensure that all degenerate measurements are inferior by choosing $r_x$ such that the linear combination of the vectors with $r_x$ is the null vector. It is possible as the combination coefficients in this case are all positive. If a linear combination is zero in a given frame, it is zero in all frames, including the one where the components of the vectors are $\mu_{xy}$, therefore $\sum_x r_x\mu_{xy}=\sum_x w_{xy}=0$, that is, the sum of the elements in all columns of $w$ is zero, therefore, the maximum of $W$ is given by genuine two-outcome measurements, indeed. However, this is a sufficient condition and not a necessary one, and it is not even sure that this is the best choice for the self-test. Therefore, there still remains some freedom to choose $r_x$. At the same time, if the null vector cannot be given as a linear combination of vectors $\vec m_x$ with strictly positive coefficients, we do not know if it can always be achieved with an appropriate choice of $r_x$ that the degenerate measurements are inferior. However, even in that case it is not necessary to double the number of rows of $w$ and the number of states to be self-tested: it is always sufficient to add one single state. Taking any positive numbers $r'_x$ and adding the vector $-\sum_x r'_x\vec m_x$ normalized, one gets a set of $M_m+1$ vectors from which one can combine the null vector with positive coefficients.

Another requirement for $w$ to be appropriate for the self-test is that $\gamma_{yy'}=0$ must be the only optimum solution. This is true if the Hessian is negative definite. It has been shown in Appendix~D of Ref.~\cite{drotos} that negative semidefiniteness always holds, so that it is sufficient that the rank is maximal, that is, $M_v(M_v-1)/2$. The rank is the same as the rank of the matrix whose elements are $t_{x\alpha}=r_x^2\mu_{xy}\mu_{xy'}$, where the pair of indices $yy'$ has been replaced by the single index $\alpha$ on the left-hand side. This matrix has $M_m$ rows and $M_v(M_v-1)/2$ columns, and due to Eq.~(\ref{eq:rmuanysize}) its rows are linearly dependent. Therefore, the Hessian may only have maximal rank if $M_m\geq M_v(M_v-1)/2+1$ holds. The optimum will not be unique otherwise.

Note that this condition implies, e.g., that self-testing a configuration of three states with non-coplanar Bloch vectors is only possible if we introduce an additional, auxiliary preparation. However, even if the condition is satisfied, the rank can still be lower than maximum, but only if the states to be self-tested and the parameters $r_x$ are chosen in a very special way. When this is the case, one may choose a different set of $r_x$, or, as a last resort, add some additional state to the set to be self-tested.

In Appendix~\ref{sec:example}, we provide an example that illustrates how the algorithm presented in this section to determine an appropriate matrix $w$ and vectors $\vec v_y$ works. For this purpose, we choose three coplanar vectors $\vec m_x$.

\section{Conclusion}\label{sec:conclusion}

We have addressed the issue of certifying non-projective measurements on qubits. For this purpose, we have considered a PM scenario and aimed at designing self-tests within this framework. In particular, we have adopted the scheme proposed by Tavakoli et~al. in Ref.~\cite{tavakoli2020self} and exemplified how to self-test, by means of linear witnesses, (i) any four- or three-outcome extremal POVM with the help of three and two projective measurements, respectively, and also (ii) any number of states with the help of as many projective measurements as the dimensionality of the space spanned by the corresponding Bloch vectors. These setups are conjectured to be minimal in terms of the number of preparations and projective measurements and represent a considerable improvement in this regard compared to the more general method of Navascu\'es et~al.~\cite{navascues2023}. The results of that work count six projective measurements for the self-testing of four-outcome qubit POVMs. In Appendix~\ref{sec:4x6}, we heuristically construct a witness to implement this self-test.

We emphasize here that utilizing a nonlinear witness could not lead to any advantage, since our problem is convex, according to allowing for shared randomness. For non-convex problems, nonlinear witnesses can be useful; see, e.g., Ref.~\cite{bowles2014} for how to generate them.

As any self-test in a PM scenario, our methods are semi-DI; as explained in the Introduction, it is impossible to design fully DI certification schemes within the PM framework. The physical assumption is a dimension bound in our case. In particular, as mentioned, we restrict the preparations to the qubit space. The work of Navascu\'es et~al.~\cite{navascues2023}, addressing Hilbert spaces of any dimension, motivates us to pose the open problem of finding PM setups and corresponding self-testing schemes involving preparations in higher dimensions with as few preparations and projective measurements as possible. Different kinds of physical assumptions also form a natural direction for generalization. However, perhaps the most intriguing question is whether we have really found the minimal setups or whether further simplification is possible.

It is certainly true that the number of ``auxiliary'' measurements can be further decreased by considering non-projective POVMs for this purpose~\cite{pauwels2022}. Notwithstanding, we can actually show that two projective measurements are not sufficient to self-test an arbitrary set of prepared states. In particular, according to Section~\ref{sec:general}, one must be able to express the Bloch vectros of all of the states to be self-tested as linear combinations of the Bloch vectors characterizing the measurements. If we have only two of the latter, their linear combinations will span a two-dimensional plane, not allowing to self-test configurations of states whose Bloch vectors span the three-dimensional space. Note that the argumentation of Section~\ref{sec:general} relies on a linear witness, but we have pointed out that no advantage can be provided by nonlinear witnesses in our scenario.

Whether or not all of our setups are truly minimal, we believe that they may contribute to an efficient implementation of the certification of non-projective measurements or configurations of states relevant for practice. To achieve this, it is necessary to address the effect of noise.

While noise inhibits achieving a self-test in a strict sense in any real-world setup, a witness designed for self-testing can be used to certify, up to sampling uncertainty, a single but well-defined qualitative property that the targeted measurement or configuration possesses but some restricted set of all possible measurements or configurations does not, as in, e.g., Ref.~\cite{tavakoli2020self}. This is because the maximum witness value attainable within such a restricted set will be strictly lower than the ideal witness value. We exemplify this approach in Appendix~\ref{sec:implementing} through the certification of the non-coplanar nature of Bloch vectors characterizing an actual configuration of states. For this purpose, we consider a witness from the family of Eq.~(\ref{eq:w4X3}) that is constructed to self-test a specific configuration that we call umbrella-like.

To illustrate that this methodology is not just a far-fetched theoretical one, we have implemented this setup in the form of quantum circuits on publicly available quantum processors. By our experimental results, presented in the same Appendix, we have successfully certified the non-coplanar nature of actual Bloch vectors and thereby the existence of a complex Hilbert space in these experiments under the assumption of a dimensionality of two. Details on the implementation are given in Appendix~\ref{sec:qcircuit}.

As a future research direction, we are also aiming at robust self-tests in the traditional sense \cite{mckague2012}. Ref.~\cite{navascues2023} has fully characterized robust self-testing in the dimension-bounded prepare-and-measure scenario in terms of upper bounds on distances from the target; however, the scaling of these bounds makes them impractical, so that determining tighter bounds would be essential. Once semi-DI certification becomes established in real-world scenarios, it may open the avenue to the anticipated quantum technological applications.

\begin{acknowledgments}
    

We acknowledge the support of the EU (QuantERA eDICT, CHIST-ERA MoDIC) and the National Research, Development and Innovation Office NKFIH (Grants No.~K145927, No.~2019-2.1.7-ERA-NET-2020-00003 and 2023-1.2.1-ERA\_NET-2023-00009).

\end{acknowledgments}

\section*{Declarations}

\subsection*{Data availability statement}

Raw experimental counts in support of Table~\ref{tab_exp} and Supplementary Table~1 are available at \url{https://doi.org/10.5281/zenodo.10723401}.

\subsection*{Code availability statement}

A code implementing the prepare-and-measure scenario presented in Section~\ref{sec:implementing} on quantum processors according to Appendix~\ref{sec:qcircuit} is available at \url{https://doi.org/10.5281/zenodo.10723275}.

\appendix
\section{Example for self-testing three states with coplanar Bloch vectors}\label{sec:example}

Here, we will apply the algorithm presented in Section~\ref{sec:anynumber} for self-testing three specific states with coplanar Bloch vectors $\vec m_x$. In particular, we consider the following Bloch vectors:
\begin{align}
\vec m_1 &= (1,0,0),\nonumber\\
\vec m_2 &= (1/2,0,\sqrt{3}/2),\nonumber\\
\vec m_3 &= (-\sqrt{3}/2,0,-1/2).
\label{eq:examplem}
\end{align}

The first step is to find coefficients $r_x$ such that the linear combination of the vectors $\vec m_x$ with $r_x$ is the null vector. In the case of three vectors as here, these coefficients are unique up to an overall constant multiplier. Let us choose them as: $r_1 = r_2 = 1$, $r_3 = \sqrt{3}$. Note that all are positive as required.

We now compute the matrix to be diagonalized:
\begin{equation}
\sum_{x=1}^3r_x\vec m_x\otimes\vec m_x=
  \left(\begin{array}{ccc}
  \frac{5+3\sqrt{3}}{4}&0&\frac{3+\sqrt{3}}{4}\\
  0&0&0\\
  \frac{3+\sqrt{3}}{4}&0&\frac{3+\sqrt{3}}{4}\\  	
  \end{array}\right).
\label{eq:examplematrix}
\end{equation}
This matrix has three eigenvalues: $(3+2\sqrt{3})/2$, $1/2$, and $0$. The corresponding normalized eigenvectors will be the Bloch vectors of the measurements. However, since the third eigenvector is $(0,1,0)$, perpendicular to the plane where the Bloch vectors $\vec m_x$ lie, it can be omitted, so that two measurement settings are sufficient, as expected. The remaining normalized eigenvectors, giving the corresponding Bloch vectors, are orthogonal to each other and are given as follows:
\begin{align}
\vec v_1 &= (\sqrt{3}/2,0,1/2),\nonumber\\
\vec v_2 &= (1/2,0,-\sqrt{3}/2).
\label{eq:examplev}
\end{align}

The coordinates of the vectors $\vec m_x$ in the coordinate system defined by the vectors $\vec v_y$, multiplied by the coefficient $r_x$, give the witness matrix $w$ as:
\begin{equation}
w=\left(\begin{array}{rr}
  \sqrt{3}/2&1/2\\
  \sqrt{3}/2&-1/2\\
  -\sqrt{3}&0\\
  \end{array}\right).
\label{eq:examplew}
\end{equation}
The sum of the elements in each column is zero by construction.

\section{Self-testing four-outcome POVMs with six projective measurements}\label{sec:4x6}

Consider an $O=4$-outcome POVM to be self-tested. Then the number of input values for Alice and the number of states to be prepared is $M_m=O=4$. Let the number of binary measurement settings for Bob be $M_v=6$. Instead of $y$, let the first $M_v$ input values of Bob be labeled by a pair of indices $ij$, where $1\le i<j\le 4$. Let us consider the following matrix for the self-test of the four states:
\begin{equation}
w_{x,ij}=F_{ij}(\delta_{xi}-\delta_{xj})
\label{eq:wmat_4x6}
\end{equation}
with $F_{ij} > 0$. As $\sum_x W_{x,ij}=0$ for any $ij$, no degenerate measurement setting can be more advantageous than the best genuine two-outcome measurement. Therefore, according to Eq.~(\ref{eq:Wfor3X4}), $W$ may be written as
\begin{equation}
W=\sum_{x=1}^{4}\sum_{i<j=1}^{4} w_{x,ij}\vec m_x\cdot\vec v_{ij},
\label{eq:Wfor4x6}
\end{equation}
which, using Eq.~(\ref{eq:wmat_4x6}), can be rewritten as:
\begin{equation}
W=\sum_{i<j=1}^{4} F_{ij}(\vec m_i-\vec m_j)\cdot\vec v_{ij}.
\label{eq:Wfor4x6s}
\end{equation}
This is the expression to be maximized in terms of the unit vectors $\vec m_x$ and $\vec v_{ij}$.

As the number of binary measurements is larger than three (the dimensionality of the Euclidean space), the method based on partial derivatives we applied in the main text is inappropriate in this case. If the optimal measurement settings are chosen in Eq.~(\ref{eq:Wfor4x6s}), the expression to be maximized in terms of the Bloch vectors of the states $\vec m_x$ is:
\begin{equation}
Q_m=\sum_{i<j=1}^{4} F_{ij}|\vec m_i-\vec m_j|.
\label{eq:Qfor4x6m}
\end{equation}
This is analogous to the case when the dependence on $\vec v_y$ is kept by choosing the optimal $\vec m_x$ vectors, like in Eq.~(\ref{eq:Qv}).

Actually, our aim is not to find the optimal $\vec m_x$ vectors for given parameters $F_{ij}$, but to find $F_{ij}$ such that Eq.~(\ref{eq:Qfor4x6m}) is maximal with the predefined set of vectors (which are anti-aligned with the vectors defining the extremal POVM to be self-tested).

It is not difficult to see that Eq.~(\ref{eq:Qfor4x6m}) corresponds to minus one times the potential energy of a simple mechanical system. The $\vec m_x$ vectors are the positions of point objects whose motion is confined to the surface of a unit sphere ($\vec m_x$ are normalized), with a constant $F_{ij}$ repulsive force acting between them (we may define $F_{ji}\equiv F_{ij}$ for $i<j$).

The potential energy is minimal and, consequently, Eq.~(\ref{eq:Qfor4x6m}) is maximal where the system is in a stable equilibrium. This happens when the resulting force acting on any of the objects points outwards, in the direction of the radius of the unit sphere. This condition can be expressed as:
\begin{equation}
\sum_{j\neq i} F_{ij}\frac{\vec m_i-\vec m_j}{|\vec m_i-\vec m_j|}=\tau_i\vec m_i,
\label{eq:equilib}
\end{equation}
where we have defined $F_{ji}\equiv F_{ij}$ for $i<j$.

We obtain the value of $\tau_i$ by taking the scalar product of both sides of the equation with $\vec m_i$:
\begin{equation}
\tau_i=\sum_{j\neq i}F_{ij}\frac{1-\vec m_i\cdot\vec m_j}{|\vec m_i-\vec m_j|}.
\label{eq:taui}
\end{equation}
As $\tau_i$ is always positive, the condition of the stability of the equilibrium is satisfied. This is not surprising: the radial component of a repulsive force between objects on the surface of a sphere obviously points outwards.

If we substitute $\tau_i$ from Eq.~(\ref{eq:taui}) into Eq.~(\ref{eq:equilib}), a simple calculation leads to:
\begin{equation}
\sum_{j\neq i}\frac{F_{ij}}{|\vec m_i-\vec m_j|}\vec m_j-\left(\vec m_i\cdot\sum_{j\neq i}\frac{F_{ij}}{|\vec m_i-\vec m_j|}\vec m_j\right )\vec m_i=0.
\label{eq:zersum}
\end{equation}
Notice that the left-hand side of the expression above is a linear combination of the $\vec m_x$ vectors, which gives the zero vector. As the $\vec m_x$ vectors are minus one times the Bloch vectors $\vec n_x$ of the POVM to be self-tested [see Eq.~(\ref{eq:povmgen})], they satisfy $\sum_x\lambda_x\vec m_x=0$. Moreover, if the POVM is extremal, all linear combinations of $\vec n_x$ and, consequently, also of $\vec m_x$ that give zero are proportional to each other. Therefore, the coefficients in Eq.~(\ref{eq:zersum}) for each $i$ must be proportional to $\lambda_j$, from which it follows, considering the first sum, that
\begin{equation}
\frac{F_{ij}}{|\vec m_i-\vec m_j|}=c_i\lambda_j
\label{eq:coeffs}
\end{equation}
with $c_i>0$, for $j\neq i$. By substituting this into the second sum in Eq.~(\ref{eq:zersum}), one finds that the coefficient of $\vec m_i$ is $c_i\lambda_i$, that is, it does have the appropriate value. Moreover, as $F_{ji}=F_{ij}$, $c_i$ must be proportional to $\lambda_i$. As an overall constant is irrelevant, we may choose $c_i=\lambda_i$. Therefore, the choice for $F_{ij}$ to get an appropriate witness matrix specified in Eq.~(\ref{eq:wmat_4x6}) for the self-test of an extremal POVM is, according to Eq.~(\ref{eq:coeffs}) and $\vec m_x=-\vec n_x$:
\begin{equation}
F_{ij}=\lambda_i\lambda_j|\vec n_i-\vec n_j|.
\label{eq:finalchoice}
\end{equation}

This method may also be used to self-test three-outcome extremal POVMs using three projective measurements (that is, $M_v=3$). It can also be extended to self-test vectors from which the null vector may be combined with not all coefficients positive. Only the signs of the appropriate rows of the $w$ matrix have to be changed. 

We refer the reader to a more general method introduced in Ref.~\cite{navascues2023} which allows one to self-test arbitrary ensembles of pure quantum states and arbitrary extremal $D\ge 2$-dimensional quantum measurements in prepare-and-measure scenarios. The method presented in the Appendix is a special case of the one described there.

\section{Certifying complex qubit state space and an implementation on quantum processors}\label{sec:implementing}

For illustrative purposes, we pick a one-parameter family from the witness matrix~(\ref{eq:w4X3}), the so-called umbrella-like construction; see the matrix~(\ref{eq:w4X3c}) parameterized by $c\in [0,3]$ below. According to the construction of the witness matrix~(\ref{eq:w4X3}), the maximum value of the witness attainable by qubit preparations is $2$ irrespective of $c$. However, to obtain this value for $0<c<3$, both the preparations and the measurements must span the complex qubit space, as shown by the self-testing argument in Sec.~\ref{sec:four-outcome}. That is, coplanar arrangements of the Bloch vectors are not sufficient to achieve the value of $2$. In this section, we provide actual bounds on the maximum value that can be attained with real qubit preparations for given $c$ values. In addition, we implement the PM setup with quantum circuits for specific values of $c$, and certify, based on the umbrella-like linear witness, that preparations with coplanar Bloch vectors are not sufficient to achieve the experimentally measured witness values in a wide range of $c$.

\subsection{The umbrella-like construction}\label{sec:umbrella}

We define the parameters in the coefficient matrix~(\ref{eq:w4X3}) by setting
\begin{align}
	p_{1} &= \frac{c}{\sqrt{3+c^2}} , \nonumber \\
	p_{2} &= p_{3} = p_{4} = \frac{1}{\sqrt{3+c^2}} , \nonumber \\
	q_{1} &= q_{2} = q_{3} = \frac{1}{\sqrt{3}},
 \label{eq:pq}
\end{align}
where $c \in [0,3]$. This way we obtain the witness matrix 
\begin{equation}
w=\frac{1}{\sqrt{9+3c^2}}\left(\begin{array}{ccc}
  \hphantom{-}c&\hphantom{-}c&\hphantom{-}c\\
  \hphantom{-}1&-1&-1\\
  -1&\hphantom{-}1&-1\\  	
  -1&-1&\hphantom{-}1\\
  \end{array}\right).
\label{eq:w4X3c}
\end{equation}

On the other hand, let us define the following unit Bloch vectors:
\begin{align}\label{eq:mx}
	\vec{m}_{1} &= \left( 0, 0, -1 \right) , \nonumber \\
	\vec{m}_{2} &= \left( -\frac{\sqrt{9-c^2}}{3}, 0, \frac{c}{3} \right) , \nonumber \\
	\vec{m}_{3} &= \left( \frac{\sqrt{9-c^2}}{6}, \frac{\sqrt{9-c^2}}{2\sqrt{3}}, \frac{c}{3} \right) , \nonumber \\
	\vec{m}_{4} &= \left( \frac{\sqrt{9-c^2}}{6}, -\frac{\sqrt{9-c^2}}{2\sqrt{3}}, \frac{c}{3} \right).
\end{align}
We call this set of Bloch vectors umbrella-like. It has the property that $\vec m_1\cdot\vec m_x$ are the same for $x=2,3,4$ and $\vec m_x\cdot\vec m_{x'}$ are the same for all other distinct pairs $(x,x')\in\{(2,3),(2,4),(3,4)\}$. 

It can be shown that the results presented in Sec.~\ref{sec:four-outcome} imply that the maximum value $W(c) = 2$ of the witness is achieved if Alice's preparations are given by Bloch vectors $\vec{m}_x$ for $x \in \{1,2,3,4\}$ in~(\ref{eq:mx}) and Bob's projective measurements are
\begin{align}\label{eq:vy}
	\vec{v}_{1} &= r\left( -2\sqrt{(9-c^2)}, 0, -4c\right)\,, \nonumber \\
	\vec{v}_{2} &= r\left( \sqrt{9-c^2}, \sqrt{3(9-c^2)}, -4c \right)\,, \nonumber \\
	\vec{v}_{3} &= r\left( \sqrt{9-c^2}, -\sqrt{3(9-c^2)}, -4c\right),
\end{align}
where $r=1/(2\sqrt{9+3c^2})$ is a normalization constant. Moreover, the above set of vectors is unique up to a rotation and a reflection through the origin. Note that apart from $c=0$ and $c=3$, both the preparation and measurement unit Bloch vectors span the full three-dimensional space. In fact, for $c=3$ all the Bloch vectors are either aligned or anti-aligned, and the witness value of $2$ can be obtained simply by preparing classical bits.

\subsection{Certifying complex qubit preparations}

Recall that the witness matrix~(\ref{eq:w4X3c}) is constructed such that the maximum witness with preparations of complex qubits is $2$ for all $c$. Our aim here is to compute bounds on the witness using real qubit preparations. That is, we consider the witness matrix~(\ref{eq:w4X3c}) for a given $c$ and we calculate the maximum value of the witness attainable with real qubit preparations and real-valued measurements. Let us call this bound by $W_{\mathbb{R}^2}(c)$ for a given $c$.

We provide a family of preparations below that appears to saturate the bound for real qubits. Actually, there are two separate families, one for $c\le1$ and an other for $c\ge1$. When $c=1$, the two families are the same up to a relabeling of the inputs $x$. It turns out that $W_{\mathbb{R}^2}(c) < 2$ for any $0< c <3$ according to these families. Hence, our witness parameterized by $c$ enables us to differentiate between real and complex two-level systems across the entire range of $0< c <3$.

For $0\le c\le 1$, the following co-planar preparations are conjectured to reach the real qubit bound $W_{\mathbb{R}^2}(c)$:
\begin{align}
\vec m_1&=(1,0,0),\nonumber\\
\vec m_2&=(\cos\alpha(c),0,\sin\alpha(c)),\nonumber\\
\vec m_3&=(\cos\alpha(c),0,-\sin\alpha(c)),\nonumber\\
\vec m_4&=(1,0,0),
\label{eq:bloch_c_small}
\end{align}
where the angle $\alpha(c)$ is given by
\begin{equation}
\alpha(c) = 2\arctan{\sqrt{f(c)}}
\label{eq:angle_c_small}
\end{equation}
with
\begin{equation}
f(c)=\frac{c^2 - 2 c + 25}{-c^2 + 2 c + 15} + 4\sqrt{5}\sqrt{\frac{c^2 - 2 c + 5}{(c^2 - 2 c - 15)^2}}.    
\end{equation}
For $1\le c\le3$, on the other hand, the following set of preparations is conjectured to reach the real qubit bound $W_{\mathbb{R}^2}(c)$ of the witness:
\begin{align}
\vec m_1&=(\cos\beta(c),0,\sin\beta(c)),\nonumber\\
\vec m_2&=(\cos\gamma(c),0,-\sin\gamma(c)),\nonumber\\
\vec m_3&=(1,0,0),\nonumber\\
\vec m_4&=(1,0,0),
\label{eq:bloch_c_large}
\end{align}
where 
\begin{align}
\beta(c)&= \frac{\pi}{2} + \arctan\frac{7c^2 - 3}{\sqrt{-9 + 82c^2 - 9c^4}}, \nonumber\\
\gamma(c)&= \frac{\pi}{2} - \arctan\frac{3c^2 - 7}{\sqrt{-9 + 82c^2 - 9c^4}}.
\label{eq:angles_c_large}
\end{align}
For both families, any rotation and a reflection through the origin is permitted, \emph{and} also any relabeling involving the indices $x\in\{2,3,4\}$ but not $x=1$. However, for $c=1$, $\alpha(c)=\beta(c)=\gamma(c)$, and the two families coincide up to a relabeling involving $x=1$.

In geometric terms, there are two Bloch vectors that coincide for any $c$. For $c<1$, $\vec m_1$ must be among them, and for $c>1$, $\vec m_1$ must not be among them, while both possibilities are permitted for $c=1$. For $c\leq 1$, the two non-coinciding Bloch vectors are symmetric to the coinciding ones, whereas they are not for $1<c<3$. In particular, $\vec m_1$ forms a greater angle with the coinciding vectors than the other non-coinciding one. In the limit of $c=3$, $\vec m_1$ becomes opposite to the rest which all coincide, and we thus recover the ideal configuration which is also classical.

\begin{center}
\begin{figure*}[t!]
\includegraphics[trim=5 0 0 0,width=0.75\textwidth]{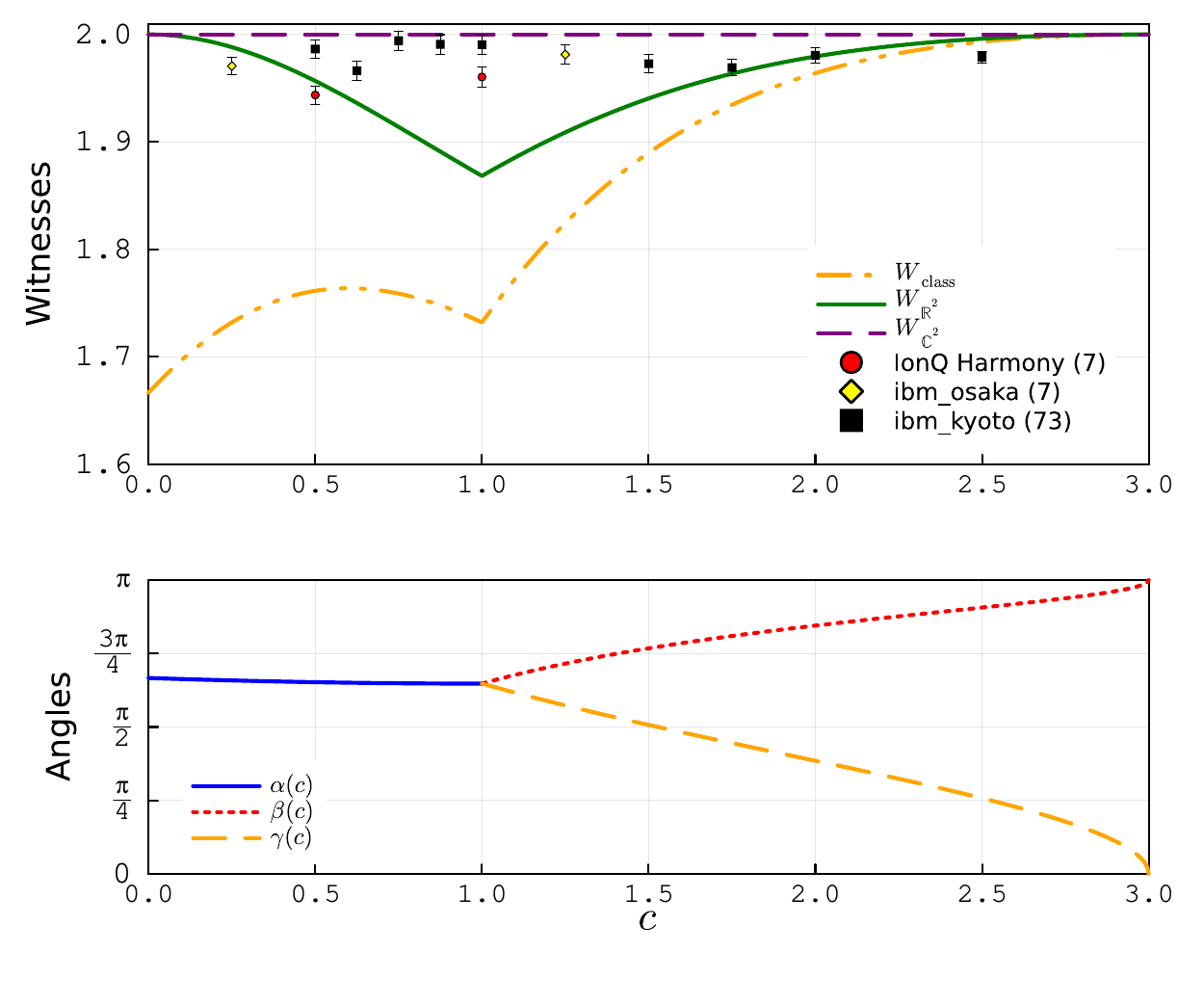}
\caption{The prepare-and-measure witness $W(c)$ defined through the matrix~(\ref{eq:w4X3c}) as a function of $c\in[0,3]$. The upper part of the figure presents the maximum witnesses attainable using classical strategies ($W_\text{class}$), real qubits ($W_{\mathbb{R}^2}$), and complex qubits ($W_{\mathbb{C}^2}$). The best experimentally measured values are also displayed for specific values of $c$, separately for IonQ and IBM machines. The error bars correspond to the estimated standard deviation associated with the finite sample size (see Appendix~\ref{sec:qcircuit}). The lower part of the figure shows the optimal angles~(\ref{eq:angle_c_small}) and~(\ref{eq:angles_c_large}) of the Bloch vectors corresponding to the maximum value of the real-qubit case for $c\le 1$ and $c\ge 1$, respectively.}
\label{fig:Wc}
\end{figure*}
\end{center}

All this is illustrated in Fig.~\ref{fig:Wc}, where witness values $W(c)$ are also displayed as a function of $c$ within the entire range $0\le c\le 3$. In particular, we show the maximum classical value $W_\text{class}(c)$ of the witness (see below) and the maximum real qubit witness $W_{\mathbb{R}^2}(c)$ in comparison with the maximum $W_{\mathbb{C}^2}(c)\equiv Q$ of the complex qubit witness; note that $W_{\mathbb{C}^2}(c)\equiv Q=2$ in the entire range $0\le c\le 3$. The range $0<c<3$ in the figure demonstrates the distinctions between the classical bit, real qubit, and complex qubit cases. Moreover, it provides a means to certify the existence of complex qubit Hilbert spaces.

The maximum value $W_\text{class}(c)$ attainable by classical bits is straightforward to determine and is obtained as
\begin{equation}
W_\text{class}(c) =
\begin{cases}
    (c+5)\frac{1}{\sqrt{3(3+c^2)}} & \text{if $c\le 1$},\\
    (c+1)\sqrt{\frac{3}{3+c^2}} & \text{if $c\ge 1$}.
\end{cases}
\label{eq:wclass}
\end{equation}

On the other hand, we compute $W_{\mathbb{R}^2}(c)$ for Fig.~\ref{fig:Wc} directly from the Bloch vectors specified by~(\ref{eq:bloch_c_small}-\ref{eq:angles_c_large}) by assuming corresponding optimal measurements in Eq.~(\ref{eq:Wfor3X4}) which then simplifies to the form $\sum_x\|\sum_y w_{xy}\vec m_x\|$, cf. Eq.~(\ref{eq:optimv}); the coefficients $w_{xy}$ are taken from (\ref{eq:w4X3c}). One may ask if the analytic formulae~(\ref{eq:bloch_c_small}-\ref{eq:angles_c_large}) merely define some generic lower bound on the exact value of $W_{\mathbb{R}^2}(c)$ or the bound is tight. To address this question, we numerically upper-bound the expression~(\ref{eq:Wfor3X4}), i.e., $\sum_{x,y} w_{x,y}\vec m_x\cdot\vec v_y$, with assuming $\vec m_x$ and $\vec v_y$ to be generic coplanar Bloch vectors. The expression is a quadratic function of the parameters defining the Bloch vectors, featuring quadratic constraints $\|\vec m_x\|\le 1$ and $\|\vec v_y\|\le 1$. To determine an upper bound $W^{(\mathrm{U})}_{\mathbb{R}^2}(c)$, we turned to the semidefinite programming approach of Lasserre~\cite{lasserre2001} which introduces moment relaxations and is applied, e.g., in Refs.~\cite{brunner2008PRL,brunner2012PRL}. In particular, we used the corresponding module built in the YALMIP package~\cite{Lofberg2004} with second-order relaxation. We considered several samples of $c$ and found in each case that the numerical upper bound $W^{(\mathrm{U})}_{\mathbb{R}^2}(c)$ obtained for real qubit preparations is saturated with our analytic lower bound up to high numerical precision. In particular, the two bounds always coincide to 7 decimals at least.
We note that for such a problem, the finite-dimensional semidefinite programming techniques introduced in Refs.~\cite{navascues2014PRX,navascues2015PRL,navascues2015PRA,tavakoli2019PRL} could also be applied.

After determining $W_{\mathbb{R}^2}(c)$ to high precision, we may ask if it is violated in certain experimental settings. In particular, we implemented the PM setup for the umbrella-like configuration as quantum circuits on quantum processors of IBM~\cite{ibmq} and IonQ~\cite{ionq}; details of the implementation can be found in Appendix~\ref{sec:qcircuit}.

We considered various values of $c$ and repeated most experiments several times to study reproducibility. We present the best experimentally measured witness values for each considered $c$, separately for IBM and IonQ, in Table~\ref{tab_exp} and include them in Fig.~\ref{fig:Wc} as well. The experimental data, in comparison with the numerical upper bounds $W^{(\mathrm{U})}_{\mathbb{R}^2}(c)$, show convincingly that we can certify complex qubit preparations in the PM setup by making use of the umbrella-like witness defined through~(\ref{eq:w4X3c}).

However, not all values of $c$ are suitable for this purpose; for $c=2.5$, we could not even certify qubits, as the best experimental witness value can be achieved with classical bits as well. Furthermore, the IBM processors do not appear to be very stable: as an extreme example, the experimental witness value for $c=0.625$ dropped from $1.95$ to $1.55$ in 25 hours on qubit N\textsuperscript{o} 7 of ibm\_osaka, and calibration data indicating the quality drop were not released by the time of the execution of the first affected experiment. Results for the witness and its estimated standard deviation from all of our experiments performed on quantum processors, along with metadata including calibration data for the machine associated with the individual experiments, are provided in Supplementary Table 1; the corresponding raw outcome counts can also be accessed~\cite{raw}.

\begin{table}
\small
\begin{tabular}{l r r c r} \hline
$c$ & $W$ & $\sigma_W$ & Machine (N\textsuperscript{o} qubit) & $W^{(\mathrm{U})}_{\mathbb{R}^2}(c)$ \\ \hline
$0.25$ & $1.9706$ & $0.0081$ & ibm\_osaka (7) & $1.9881$ \\
$0.5$ & $1.9436$ & $0.0086$ & IonQ Harmony (7) & $1.9567$ \\
$0.5$ & $1.9865$ & $0.0085$ & ibm\_kyoto (73) & $1.9567$ \\
$0.625$ & $1.9663$ & $0.0088$ & ibm\_kyoto (73) & $1.9362$ \\
$0.75$ & $1.9943$ & $0.0089$ & ibm\_kyoto (73) & $1.9139$ \\
$0.875$ & $1.9908$ & $0.0090$ & ibm\_kyoto (73) & $1.8910$ \\
$1$ & $1.9604$ & $0.0091$ & IonQ Harmony (7) & $1.8683$ \\
$1$ & $1.9905$ & $0.0090$ & ibm\_kyoto (73) & $1.8683$ \\
$1.25$ & $1.9814$ & $0.0089$ & ibm\_osaka (7) & $1.9089$ \\
$1.5$ & $1.9728$ & $0.0085$ & ibm\_kyoto (73) & $1.9404$ \\
$1.75$ & $1.9692$ & $0.0079$ & ibm\_kyoto (73) & $1.9635$ \\
$2$ & $1.9805$ & $0.0070$ & ibm\_kyoto (73) & $1.9795$ \\
$2.5$ & $1.9786$ & $0.0050$ & ibm\_kyoto (73) & $1.9960$ \\
\hline
\end{tabular}
\caption{\label{tab_exp}The best experimentally measured witness $W$ for different values of $c$, separately for IonQ and IBM machines, in comparison with the corresponding numerical upper bounds $W^{(\mathrm{U})}_{\mathbb{R}^2}(c)$. The standard deviation $\sigma_W$ is associated with the finite number of shots ($8192$ in all cases) and has been estimated from the experimentally obtained probabilities (see Appendix~\ref{sec:qcircuit}).}
\end{table}

Note that a similar distinguishability of real and complex Hilbert spaces of given dimensionality can also be considered in the Bell nonlocality setup. In this scenario, one compares the maximum quantum violation of Bell inequalities achieved using real qubit component spaces with that achieved using complex qubit component spaces. See, e.g., Ref.~\cite{pal2008PRA} and Appendix~B in Ref.~\cite{gisin_questions}. In addition, the more demanding task of distinguishing real and complex Hilbert spaces without fixing their dimension has been achieved in the network configuration~\cite{renou2021,bednorz2022,yao2024,batle2024} along with experimental validations~\cite{chen2022PRL,li2022PRL}.

\section{Details of implementing the PM setup on a quantum processor}\label{sec:qcircuit}

Our task is to implement the preparations and measurements of a PM scenario as a set of quantum circuits. These can then be submitted to quantum processors.

From the computational basis state $\ket{0}$, an arbitrary state $\alpha\ket{0}+\beta\ket{1}$ can be obtained by applying the unitary gate
\begin{equation}\label{eq:U_prep}
	U_\mathrm{prep} =
	\begin{pmatrix}
		\alpha & -\beta^* \\
		\beta & \alpha^*
	\end{pmatrix},
\end{equation}
where the second column follows from unitarity and is determined up to a factor of $-1$. The coefficients $\alpha$ and $\beta$ are related to the coordinates of the corresponding Bloch vector $\vec{b} = (b_1,b_2,b_3)$ as
\begin{align}
	\alpha &= \sqrt{\frac{1+b_3}{2}} , \nonumber \\
	\beta &= \frac{b_1+ib_2}{\sqrt{b_1^2+b_2^2}}\sqrt{\frac{1-b_3}{2}} ,
\end{align}
assuming $\sqrt{b_1^2+b_2^2}\neq0$ for $\beta$ (in the opposite case, $\beta=0$). For our purposes, we shall identify $\vec{b}$ with each $\vec{m}_x$, $x \in \{1,2,3,4\}$, of Eq.~(\ref{eq:mx}).

Quantum processors are designed to perform projective measurements in the computational basis. In order to emulate a projective measurement in a different basis, one can define a unitary gate by identifying the rows of the corresponding matrix $U_\mathrm{proj}$ with the complex conjugates of the desired basis states. If this unitary operator is applied to a given state, then performing a measurement in the computational basis on the new state will be equivalent to performing a measurement in the desired basis on the original state. In terms of a Bloch vector $\vec{v} = (v_1,v_2,v_3)$ characterizing the first basis state defining the desired projective measurement, $U_\mathrm{proj}$ is obtained as
\begin{equation}\label{eq:U_proj}
	U_\mathrm{proj} =
	\begin{pmatrix}
		\cos\frac{\theta}{2} & e^{-i\phi}\sin\frac{\theta}{2} \\
		\sin\frac{\theta}{2} & -e^{-i\phi}\cos\frac{\theta}{2}
	\end{pmatrix},
\end{equation}
where
\begin{align}
	\cos\frac{\theta}{2} &= \sqrt{\frac{1+v_3}{2}} , \nonumber \\
	\sin\frac{\theta}{2} &= \sqrt{\frac{1-v_3}{2}} , \nonumber \\
	e^{-i\phi} &= \frac{v_1-iv_2}{\sqrt{v_1^2+v_2^2}}
\end{align}
are defined through the polar and azimuthal angles $\theta$ and $\phi$ of $\vec{v}$, if $\sqrt{v_1^2+v_2^2}\neq0$ ($U_\mathrm{proj}=\mathbb{1}$ or $U_\mathrm{proj}=\sigma_1$ otherwise). We shall substitute $\vec{v} = \vec{v}_y$ for each $y \in \{1,2,3\}$ from Eq.~(\ref{eq:vy}).

After these considerations, both elements are given to build the quantum circuits that implement the PM setup for the umbrella-like preparations. In fact, this requires a collection of 12 different quantum circuits, each realizing one preparation (out of four) from Alice and one measurement (out of three) from Bob. The practical implementation has been carried out with Qiskit~\cite{Qiskit}.

In principle, when constructing the outcome statistics for computing the witness, one of the 12 different circuits should be chosen randomly for each shot. However, it is equivalent and much more resource efficient to perform a random but predefined number of shots consecutively with each circuit. Following Ref.~\cite{baumer2021}, we further simplify the experiments by performing the \emph{same} number, $N$, of shots with each circuit.

With this choice, the standard deviation $\sigma_W$ corresponding to the sampling distribution of the witness~(\ref{eq:witnesswithprob}) can be calculated in terms of the conditional probabilities $P(b|xy)$; the witness matrix is given by~(\ref{eq:w4X3c}) for the PM setup for the umbrella-like preparations. Each term in Eq.~(\ref{eq:witnesswithprob}) has a binomial sampling distribution, and the perfect anticorrelation between the opposing outcomes of a projective measurement has to be taken into account for the propagation of the error through the summation.

As in Ref.~\cite{baumer2021}, we estimate the standard deviation characterizing each experiment by substituting the experimentally obtained relative frequencies in place of the conditional probabilities. We furthermore note that the standard deviation corresponding to ideal umbrella-like preparations and associated measurements can be analytically computed, as a function of $c$, as
\begin{equation}
	\sigma_W = \frac{1}{3+c^2}\sqrt{\frac{27+42c^2-5c^4}{6N}} .
\end{equation}

%

\end{document}